\documentclass[12pt]{article}

\usepackage[margin=1in]{geometry}
\usepackage{amsmath,amssymb}
\usepackage{graphicx}
\usepackage{siunitx}
\usepackage{booktabs}
\usepackage{multirow}
\usepackage{xr-hyper} 
\usepackage{hyperref}
\usepackage{cleveref}
\usepackage{xcolor}
\usepackage[super,sort&compress]{natbib}

\newcommand{\degree}{\ensuremath{{}^\circ}}

\newcommand{\safeincludegraphics}[3][]{%
  \IfFileExists{#2}{\includegraphics[#1]{#2}}{\fbox{\parbox[c][0.22\textheight][c]{0.9\linewidth}{\centering Missing figure: #3}}}%
}

\begin{document}

\title{End-to-End Quantum Key Distribution Across Hybrid Fiber and Free-Space Links with All-Optical Encoding Conversion}

\author{\begin{tabular}{c}
Khen Cohen$^{1,*, \dagger}$, Tomer Nahum$^{1,2,\dagger}$, Michael Tzukran$^{1}$,\\
Paz Or$^{3}$, Yehuda Pilnyak$^{3}$, Nitzan Livneh$^{3}$, Hagai Eisenberg$^{3,4}$, \\
Yaron Oz$^{1}$, and Haim Suchowski$^{1}$ \\
\small $^{1}$Raymond and Beverly Sackler School of Physics and Astronomy, \\
\small Faculty of Exact Sciences, Tel Aviv University, Tel Aviv, Israel\\
\small $^{2}$School of Electrical Engineering, Iby and Aladar Fleischman, \\
\small Faculty of Engineering, Tel Aviv University, Tel Aviv, Israel\\
\small $^{3}$HEQA Security, Modi’in, Israel\\
\small $^{4}$Racah Institute of Physics, Faculty of Science, \\
\small The Hebrew University of Jerusalem, Jerusalem\\
\small $^{*}$Corresponding author: \href{mailto:khencohen@mail.tau.ac.il}{khencohen@mail.tau.ac.il}\\
\small $^{\dagger}$These authors contributed equally
\end{tabular}}

\date{}
\maketitle

\begin{abstract}
Quantum key distribution (QKD) promises information-theoretically secure communication, but future networks must bridge fiber and free-space links that naturally employ different photonic encodings, namely time-bin in fiber and polarization in free space.
Here we demonstrate a complete hybrid fiber and free-space QKD link that bridges both media within a single end-to-end protocol, converting between the two encodings entirely in the optical domain.
Using the decoy-state BB84 protocol operating at \SI{1550}{nm}, we demonstrate continuous secure-key generation over a \SI{90}{m} outdoor free-space link.
The system operates across atmospheric conditions spanning more than two orders of magnitude in the refractive-index structure parameter $C_n^2$, from strong daytime turbulence to quiescent nighttime conditions, and we further validate photon-level operation over a \SI{750}{m} free-space extension.
Throughout, the link maintains a session-mean quantum bit error rate (QBER) of \SIrange{5.6}{6.8}{\percent}, well below the \SI{11}{\percent} BB84 security threshold.
The encoding conversion is performed entirely in the optical domain without measurement or state reconstruction, preserving the security assumptions of the BB84 protocol.
Consequently, the time-bin-to-polarization (T2P) and polarization-to-time-bin (P2T) converters remain part of the untrusted quantum channel rather than trusted intermediate nodes.
These results establish secure photonic encoding conversion as a practical interface between fiber and free-space quantum communication platforms, providing a building block for future quantum networks applications.
\end{abstract}

\section{Introduction}

Quantum communication, which provides information-theoretically secure key distribution based on the laws of quantum mechanics, has progressed rapidly over the past two decades \cite{bennett1984, ekert1991, gisin2002}.
Secure quantum links have been demonstrated in both terrestrial optical fiber networks and long-range free-space channels, including satellite-based systems \cite{peev2009, chen2021}.
These complementary platforms provide distinct advantages: fibers allow dense integration with existing telecommunication infrastructure, whereas free-space channels support communication over continental and global distances \cite{ursin2007, bedington2017, yin2017, liao2017}.
Future quantum communication networks are expected to combine both media to provide secure connectivity across local, metropolitan, and global scales.

However, realizing such heterogeneous quantum networks requires an interconnection between these fundamentally different media that does not measure or sample the quantum signal.
Such an interconnection extends the reach of a single free-space ground station, which is otherwise limited by geographically constrained line-of-sight requirements and atmospheric turbulence, so that one station can serve the whole surrounding area.
Since the quantum signal is never measured at the conversion points, it also removes the need to physically secure them.

A central challenge arises because different physical channels favor different photonic encoding methods.
Polarization encoding \cite{bennett1984} is particularly attractive in free-space propagation due to its resilience to atmospheric turbulence and ease of state preparation and measurement \cite{ursin2007, liao2017}.
Conversely, time-bin and phase encoding \cite{brendel1999} dominate fiber-based systems because they are largely immune to polarization drift and polarization-mode dispersion \cite{scarani2009, boaron2018}.
As a result, future hybrid quantum networks will require efficient interfaces capable of transferring quantum information between different photonic encoding methods without compromising security or performance.
While several studies have explored hybrid fiber or free-space quantum communication and photonic encoding conversion \cite{rossi2026intermodal, picciariello2025intermodal, cocchi2025time, scalcon2022cross}, an end-to-end demonstration of quantum key distribution across a hybrid fiber and free-space link with in-line encoding conversion has not yet been reported.

Here, we demonstrate a complete end-to-end hybrid quantum communication link that connects fiber and free-space channels through conversion between time-bin and polarization encodings in both directions.
The conversion is performed entirely in the optical domain, without measurement or state reconstruction, making it transparent to the BB84 security analysis.
Consequently, the T2P and P2T converters remain part of the untrusted quantum channel rather than trusted intermediate nodes, preserving the information-theoretic security of the distilled key.
Using a decoy-state BB84 QKD protocol at \SI{1550}{nm}, we establish secure key generation over outdoor links operating across diverse atmospheric conditions spanning weak to mid-strong turbulence regimes.
We demonstrate sustained end-to-end QKD operation over a \SI{90}{m} hybrid link and validate photon-level operation over a \SI{750}{m} free-space extension.
We analyze the hybrid channel under diverse environmental conditions, including atmospheric turbulence, alignment drift, and background noise.
Across these conditions, the system consistently maintains a quantum bit error rate (QBER) under \SI{7}{\percent}, well below the \SI{11}{\percent} BB84 security threshold~\cite{shor2000}, while continuously generating secure cryptographic keys.
By bridging the photonic encodings naturally preferred by fiber and free-space channels, our results establish a practical interface between terrestrial and long-range quantum communication platforms, providing a building block for future heterogeneous quantum networks.

\section{Results}
\label{sec:results}
\begin{figure}[t!]
  \centering
  \includegraphics[width=\textwidth, trim={10 75 10 75}, clip]{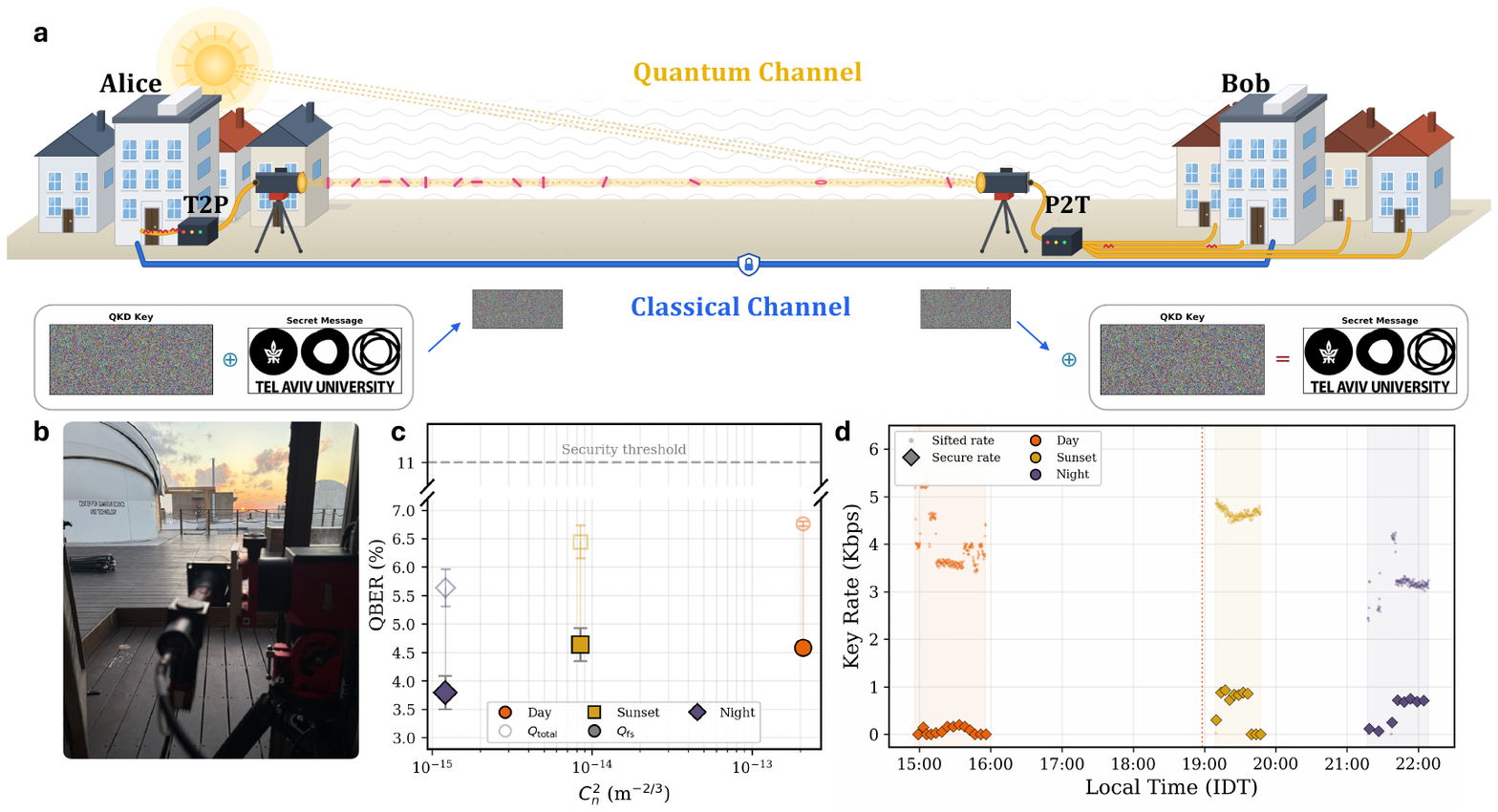}
  \caption{
    \textbf{Hybrid fiber and free-space quantum communication link and experimental performance.}
    \textbf{a}~Schematic of the hybrid quantum communication architecture. Quantum information is transmitted through a channel comprising optical fiber (time-bin encoding) and free-space propagation (polarization encoding), using T2P and P2T converters between the interfaces. Secure keys generated through the QKD protocol are used for one-time-pad encryption and decryption over an authenticated classical channel.
    \textbf{b}~Photograph of the rooftop experimental setup at Tel Aviv University.
    \textbf{c}~Session-averaged free-space (and total) QBER as a function of the estimated atmospheric turbulence strength, quantified by the refractive-index structure parameter $C_n^2$. Measurements were performed during Day, Sunset, and Night sessions spanning more than two orders of magnitude in turbulence conditions. All measurements remain well below the BB84 security threshold.
    \textbf{d}~Temporal evolution of the sifted-key rate and the secure-key rate (the latter after error-correction coding and privacy amplification) versus local time, demonstrating continuous secure-key generation under varying environmental conditions.
  }
  \label{fig:main}
\end{figure}

The hybrid quantum communication system was deployed on an urban rooftop at Tel Aviv University and operated at a wavelength of \SI{1550}{nm} (Fig.~\ref{fig:main}a).
Experiments were performed over two outdoor free-space distances, a \SI{90}{m} link and a \SI{750}{m} round-trip link to a nearby building.
The \SI{90}{m} link was evaluated across three sessions spanning a full diurnal cycle that includes Day ($15{:}00$ IDT), Sunset ($19{:}00$ IDT), and Night ($21{:}30$ IDT).
These sessions covered a wide range of atmospheric conditions.
The \SI{750}{m} link was investigated during an additional measurement session between the Day and Sunset experiments.
Throughout all measurements, environmental parameters, including temperature, humidity, dew point, and wind speed, were continuously monitored to characterize the atmospheric channel.

The transmitter (Alice) comprises a fiber-based time-bin QKD module (decoy-state BB84, developed by HEQA Security) interfaced to the outdoor free-space channel through a custom T2P converter.
Before the receiver (Bob), a complementary P2T converter converts the signal back to time-bin encoding, independently of the quantum-state analysis.
The free-space channel was implemented using a passive retroreflector architecture, allowing a hybrid fiber and free-space quantum link without the need for actively aligned free-space terminals.
This configuration provides a practical testbed for evaluating encoding conversion and hybrid quantum communication under realistic environmental conditions.
Details of the experimental implementation, optical loss budget, and noise analysis are provided in Methods and Supplementary Note~\ref{supp:loss-noise}.

\subsection{Performance of the end-to-end hybrid QKD link}
\label{sec:qkd-results}

The hybrid fiber and free-space link maintained secure quantum communication across all atmospheric conditions investigated, with QBER values consistently below the security threshold and continuous secure-key generation throughout the measurement campaign.

A total of \SI{2.88}{Mbit} (\SI{352}{KiB}) of secure key material was shared over approximately 2.5~hours of cumulative operation.
Fig.~\ref{fig:main}a demonstrates an information encryption and decryption example using the shared key generated during the experiment.
While the QBER threshold for the BB84 protocol is approximately \SI{11}{\percent}~\cite{shor2000}, the system was tuned to exchange secure keys only when the QBER is below \SI{7.5}{\percent}, keeping a stronger security margin.

Table~\ref{tab:qkd-summary} summarizes the performance achieved under the three atmospheric conditions.
The mean QBER ranged from \SI{5.64}{\percent} during the Night to \SI{6.75}{\percent} during the Day session, remaining consistently below the security threshold.
The tabulated QBER is computed over all acquired blocks, whereas the secure key is distilled only from blocks whose QBER lies below the \SI{7.5}{\percent} acceptance threshold.
Sifted key rates averaged \SIrange{3.2}{4.6}{kbps}, while the secure key rate, after error correction and privacy amplification, reached a peak instantaneous value of \SI{0.93}{kbps} during the Sunset session (Fig.~\ref{fig:main}d).
In total, 24~key blocks (8~per session) were successfully reconciled, with individual block sizes ranging from \SI{64}{kbit} to \SI{1}{Mbit}.

\begin{table}[htbp]
\centering
\caption{
  Summary of QKD performance across the three measurement sessions.
}
\label{tab:qkd-summary}
\small
\begin{tabular}{lccccc}
\toprule
Session & Duration & QBER & Sifted rate & Secure rate & Secure key \\
        & (min)    & (\%) & (kbps)      & (kbps)      & (kbit) \\
\midrule
Day (15:00--16:00)     & 60 & $6.75 \pm 0.62$ & $4.01 \pm 0.57$ & $0.067 \pm 0.078$ & $\sim$228 \\
Sunset (19:09--19:47)  & 38 & $6.45 \pm 3.56$ & $4.63 \pm 0.39$ & $0.568 \pm 0.402$ & $\sim$1{,}273 \\
Night (21:17--22:08)   & 51 & $5.64 \pm 3.96$ & $3.16 \pm 0.58$ & $0.499 \pm 0.296$ & $\sim$1{,}381 \\
\midrule
\textbf{Total}         & \textbf{149} & --- & --- & --- & $\sim$\textbf{2{,}882} \\
\bottomrule
\end{tabular}
\end{table}

The measured performance exhibits a clear dependence on environmental conditions.
The Day session showed the highest QBER and lowest secure key rate, consistent with the elevated daytime background noise and stronger near-ground turbulence from solar heating.
The Night session yielded the lowest QBER, with the Sunset session falling in between.
We attribute the low nighttime QBER primarily to the reduced background-count fraction after dark.
The Sunset session achieved the highest overall secure key rate.
The secure-key rate was not fully correlated with the QBER, because it is also affected by the hold-off period through the sifted key rate (see Discussion).

To explore the scalability of the hybrid architecture, we extended the free-space segment to \SI{750}{m}.
Photon detections were registered on both single-photon detectors, confirming operation of the hybrid quantum channel at this distance.
While optimal alignment briefly produced transmission rates approaching \SI{0.2}{kbps}, the sustained count rate remained below the threshold required for continuous QKD operation.
Compared with the \SI{90}{m} link, the \SI{750}{m} configuration incurred an additional \SI{3.8}{dB} round-trip transmission penalty (\SI{10.3}{dB} versus \SI{6.5}{dB}), primarily attributable to implementation-specific alignment and beam-expansion imperfections rather than the encoding-conversion architecture itself.
These observations indicate that the current limitation is an engineering challenge rather than a fundamental physics one, supporting the scalability of the hybrid architecture toward kilometer-scale free-space links.

\subsection{Hybrid-link performance under atmospheric turbulence}
\label{sec:turbulence}

We characterize the atmospheric channel by estimating the refractive-index structure parameter $C_n^2$, a measure of local turbulence strength, from concurrent meteorological data.
From this, we derive the Rytov variance $\sigma_R^2$, which quantifies the rapid intensity fluctuations (scintillation), the Fried parameter $r_0$, representing the spatial coherence length, and the aperture-averaged scintillation index $\sigma_I^2(D)$, which measures the intensity fluctuations that remain after spatial smoothing by the \SI{50.8}{mm} receiver aperture (the full procedure, including the Monin-Obukhov surface-layer scaling~\cite{Wyngaard1971} and the Andrews-Phillips Gaussian-beam formulation, is detailed in Supplementary Note~\ref{supp:turbulence}).

\paragraph{QBER decomposition into signal and background}
To isolate the turbulence-induced error from the background-noise floor, we decompose the measured QBER into a signal contribution, which is composed of the fiber and free-space sources of errors, and a background-noise contribution.
The background counts arise from detector dark counts and stray light, and since their state is completely random, their error probability is exactly $e_0 = 0.5$.

This background noise is measured via periodic vacuum windows, during which the transmitter sends no photons.
Denoting the total detection rate by $C_{\text{total}}$, the background rate by $C_{\text{bg}}$, and the signal rate by $C_{\text{sig}} = C_{\text{total}} - C_{\text{bg}}$, the measured QBER is a rate-weighted sum~\cite{ma2005}:
\begin{equation}
  Q_{\text{total}} = \left(Q_{\text{fib}} + Q_{\text{fs}}\right) \frac{C_{\text{sig}}}{C_{\text{total}}} + \frac{1}{2}\frac{C_{\text{bg}}}{C_{\text{total}}}.
  \label{eq:qber-hybrid-decoupled}
\end{equation}
Here, $Q_{\text{fib}}$ and $Q_{\text{fs}}$ represent the intrinsic optical errors originating from the fiber segment (e.g., polarization drifts) and the free-space segment (e.g., atmospheric turbulence), respectively.

Since the system is end-to-end secure, we cannot directly distinguish between the free-space intrinsic error $Q_{\text{fs}}$ and the fiber intrinsic error $Q_{\text{fib}}$.
To address this, we performed an additional fiber-only experiment and measured $Q_{\text{fib}}$ at six attenuation levels.
The average $Q_{\text{fib}}$ was found to be approximately \SI{1.45}{\percent}, and further details of this experiment are given in Supplementary Note~\ref{app:fiberQBEReval}.

\begin{figure}[t!]
  \centering
  \includegraphics[width=\textwidth, trim={5 150 5 130}, clip]{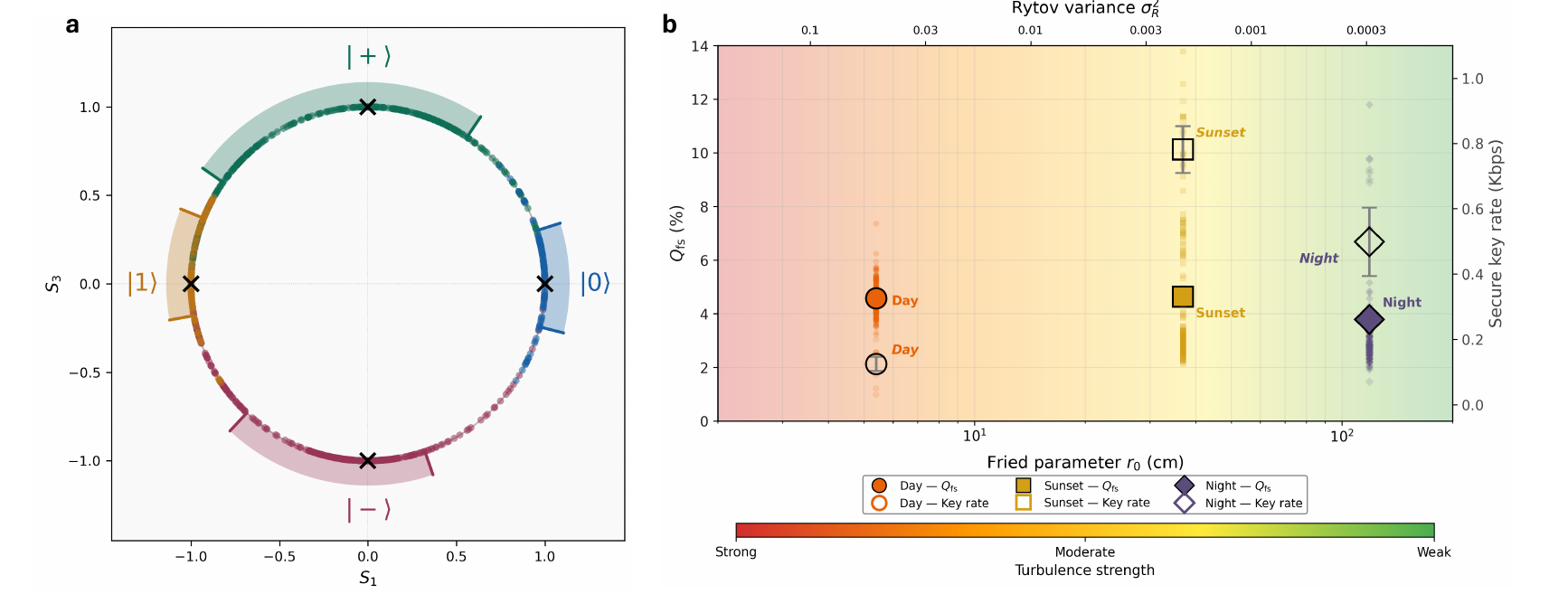}
  \caption{ \textbf{Turbulence and noise sources.}
  \textbf{a}~End-to-end quantum state evaluation for the four different states. The shaded area represents $\pm 1\sigma$ around the mean distribution.
  \textbf{b}~Turbulence measurement results. The bottom axis shows the Fried coherence diameter $r_0$ and the top axis shows the corresponding Rytov variance $\sigma_R^2$. Both are derived from the estimated $C_n^2$. The session-averaged free-space QBER values are shown as filled markers (left axis), while the secure key rate is shown by hollow markers (right axis). Error bars indicate $\pm 1 \sigma$ around the mean value. The background color represents the relative turbulence strength from strong (red) to weak (green), as quantified by the color bar.
  }
  \label{fig:turb-qkd}
\end{figure}

The system maintained stable QKD performance across atmospheric conditions spanning more than two orders of magnitude in turbulence strength, keeping the QBER well below the BB84 security threshold (Fig.~\ref{fig:main}c).
In the $xz$ projection of the Bloch sphere, the measured quantum states after transmission through the hybrid link cluster around the four ideal BB84 states generated by Alice (Fig.~\ref{fig:turb-qkd}a), showing that the transmitted qubit is preserved despite propagation through the hybrid fiber and free-space channel.
The QKD performance also tracks the atmospheric turbulence (Fig.~\ref{fig:turb-qkd}b).
The free-space QBER component $Q_{\text{fs}}$ and the secure key rate follow the Rytov variance $\sigma_R^2$ and the Fried parameter $r_0$, which quantify the strength of optical scintillation and the atmospheric coherence length, respectively.
Additional turbulence analysis is given in Supplementary Note~\ref{supp:turbulence}.

\section{Discussion}
\label{sec:discussion}

The ability to interconnect fiber and free-space quantum channels is an important requirement for extending quantum communication beyond the limits of a single transmission medium.
In the presented architecture, quantum information is transferred between time-bin and polarization encodings without measurement or state reconstruction, allowing each channel to operate in its preferred encoding while preserving end-to-end BB84 security.
Because no measurement or sampling of the quantum state occurs at either interface, the conversion is transparent to the BB84 security analysis.
Consequently, the converters remain part of the untrusted quantum channel, preserving the information-theoretic security of the distilled key.
The interface itself, rather than the individual fiber or free-space segments, is therefore the element under test, and our measurements indicate that it introduces no measurable performance penalty beyond that imposed by the atmospheric channel.

Although the QBER remained comfortably below the security threshold throughout the experiments, maximizing the secure-key rate required balancing detector dead time against the available signal-to-noise ratio.
Three independent mechanisms ultimately limited the secure-key rate.
First, the current encoding-conversion architecture introduces intrinsic optical losses, including the \SI{3}{dB} penalty associated with each conversion stage and additional losses in the free-space interface.
These losses reduce the photon flux reaching the detectors independently of atmospheric conditions.
Second, daytime operation is affected by background photons originating from scattered and reflected sunlight, increasing the detector noise floor.
Third, atmospheric turbulence introduces fluctuations in the free-space channel that reduce the fiber-coupling efficiency (a minor effect at \SI{90}{m} that grows with range) and consequently the received signal level.
While background noise and turbulence can be mitigated through nighttime operation, improved filtering, and advanced beam-control techniques, the conversion losses arise from the present converter architecture and therefore define an important target for future optimization.
Increasing the detector hold-off period suppressed noise counts and maintained low QBER, at the expense of reducing the sifted-key and secure-key rates.
This tradeoff explains the lower secure-key rate observed during the Day session despite only a modest increase in QBER.

Compared with prior hybrid and interface-level work~\cite{scalcon2022cross}, the present demonstration moves the encoding conversion out of the laboratory and into a deployed link with realistic atmospheric variability.
Although satellite-based quantum communication has demonstrated far longer transmission distances~\cite{yin2017,liao2017}, the atmospheric path here is fundamentally different, a short horizontal near-ground path through strong $C_n^2$, rather than a long slant path through predominantly weak high-altitude turbulence.
Those experiments operate in the weak-turbulence vertical-path limit, where the Fried parameter $r_0$ is comparable to or larger than the receive aperture.
Our \SI{90}{m} link sits in the same weak-fluctuation regime ($r_0 \gtrsim D$, $\sigma_R^2 \lesssim 0.05$) as a satellite link despite the strong near-ground $C_n^2$, because $\sigma_R^2 \propto C_n^2 L^{11/6}$ and the short path accumulates little scintillation.
Satellite links also rely on much larger collecting apertures to raise the receiver gain and average over turbulence, and they operate at high elevation angles where the background noise is much lower than in the urban environment of our rooftop terminal.

At \SI{90}{m}, the turbulence diagnostics (Methods) show that free-space propagation contributes little to the residual error, because beam wander, angle-of-arrival fluctuations, and aperture-averaged scintillation are all small and fewer than one coherence cell spans the receiver aperture.
The residual $Q_{\mathrm{fs}}$ is therefore not turbulence-limited but set by a turbulence-independent conversion and fiber-coupling baseline, evidenced by its weak dependence on $C_n^2$ across the two-order-of-magnitude diurnal range, with daytime background noise contributing the remainder.
Turbulence becomes significant only over longer paths because $\sigma_R^2 \propto C_n^2 L^{11/6}$, the same $C_n^2$ that is weak at \SI{90}{m} reaches the moderate regime ($\sigma_R^2 \approx 2.4$) at \SI{750}{m} and the strong-fluctuation regime at kilometer scale.
At those distances high-order wavefront distortion would dominate the coupling loss and cannot be mitigated by using a larger receive aperture or fast tip-tilt tracking, which correct only low-order errors, so higher-order adaptive optics would be needed to recover the QKD link margin.
This scaling marks where the hybrid link shifts from conversion and noise-limited to free-space distortion-limited operation.

The experiments were conducted over a single diurnal cycle, so the reported $C_n^2$-QBER relationship should be read as representative rather than statistically exhaustive, and extending the measurements across seasons and weather conditions will be important for assessing long-term stability.
The asymmetry in the reconstructed quantum state distributions (Fig.~\ref{fig:turb-qkd}a) arises primarily from residual polarization drift that accumulates after the pre-session calibration, and implementing adaptive polarization tracking during key generation is expected to reduce this imbalance and further improve the secure-key rate, particularly during the Day session, which had the lowest secure-key rate.

The demonstrated architecture provides a practical route toward heterogeneous quantum communication infrastructures that combine fiber and free-space channels within a single security-preserving architecture.
In metropolitan areas, free-space links could bridge gaps between fiber-connected nodes while maintaining compatibility with existing fiber-based QKD systems through the encoding-conversion interface demonstrated here.
Beyond terrestrial networks, the same architecture naturally extends to airborne and satellite deployments, allowing quantum links between mobile or space-based terminals and fiber-connected users without the need for direct free-space access at every endpoint.
More fundamentally, the demonstrated interface decouples the choice of photonic encoding from the transmission medium.
Fiber and free-space channels can therefore each employ the encoding best suited to their environment while operating within a single quantum communication network.

\section{Methods}
\label{sec:methods}

The hybrid link couples an unmodified fiber-based time-bin QKD system to a turbulence-resilient free-space polarization channel through a pair of all-optical encoding converters.
It comprises four stages in series: a fiber-based time-bin QKD transmitter (Alice) with a T2P converter, a folded free-space segment terminated by a passive retroreflector, a P2T converter, and the time-bin QKD receiver (Bob).
Alice and Bob are colocated and share a single fiber-to-free-space port through an optical circulator, so the free-space path is traversed in both directions.
The following subsections describe each stage.
Component-level specifications are listed in Supplementary Note~\ref{supp:experimental-details}, and the full loss budget in Supplementary Note~\ref{supp:loss-noise}.

\subsection{Decoy-state QKD platform}
\label{sec:heqa-protocol}

The endpoints are a pre-commercial fiber-based QKD platform (HEQA Security) implementing the decoy-state BB84 protocol with time-bin encoding\cite{bennett1984,hwang2003,lo2005}, used here unmodified.
Qubits are encoded in the arrival time of weak coherent pulses at \SI{1550}{nm} within a \SI{1}{ns} window: occupation of an \emph{early} or \emph{late} slot defines the $Z$-basis states $|0\rangle$ and $|1\rangle$, while their phase-controlled superposition defines the $X$-basis states $|+\rangle$ and $|-\rangle$ (\cref{fig:setup}\textbf{a}).
Three intensity levels (signal, decoy, and vacuum) implement the decoy-state method, and at the receiver, an unbalanced Mach-Zehnder interferometer and two gated InGaAs single-photon avalanche detectors map the time-bin states onto distinct detection windows for the $Z$ and $X$ bases (\cref{fig:setup}\textbf{e}).

We characterize the link by three quantities: the \emph{sifted key rate} (coincident detections in matching bases), the \emph{quantum bit error rate} (QBER, the fraction of disagreeing sifted bits), and the \emph{secure key rate} (information-theoretically secure bits after error correction and privacy amplification).
Positive secure-key generation requires $\mathrm{QBER} \lesssim \SI{11}{\percent}$.
The decoy-state secure-key analysis (GLLP framework~\cite{Gottesman2004}, with the decoy-state bounds of Ma \textit{et~al.}~\cite{ma2005}) and the classical post-processing chain (basis reconciliation, Cascade error correction, and privacy amplification) are detailed in Supplementary Note~\ref{supp:post-processing}.
The detectors operate in gated mode with a \SI{1}{ns} gate width and a session-dependent hold-off period (\SIrange{15}{30}{ns} for Day, \SI{15}{ns} for Sunset, \SI{25}{ns} for Night).

\subsection{Time-bin and polarization conversions}
\label{sec:converter}

Bridging the fiber-based time-bin QKD system to the polarization-encoded free-space channel is the central architectural element of the hybrid link.
The interface comprises a T2P converter at the transmitter and a P2T converter before the receiver (\cref{fig:setup}\textbf{a},\textbf{d}).
Each converter is a fiber-based unbalanced Mach-Zehnder interferometer whose path-length difference matches the relevant time-bin separation ($\Delta t_A = 800$\,ps at the T2P converter and $\Delta t_B \approx 500$\,ps at the P2T converter, with the deliberate mismatch exploited for polarization calibration as described below), with a $\lambda/2$ retardation in the late-bin arm, so that early/late time bins are mapped onto vertical/horizontal polarization, and vice versa.
Unlike trusted-node architectures, this conversion is performed entirely in the optical domain, without measurement or state reconstruction.
The qubit is never detected, copied, or regenerated, so no point along the link holds a classical record of the key, and no additional trust assumption is introduced.
The mapping is passive, reciprocal, and protocol-transparent.
A related conversion has been used previously to adapt a polarization-based QKD system to fiber operation~\cite{scalcon2022cross}.
Here it is applied in the reverse sense, projecting a mature time-bin platform onto a turbulence-resilient polarization channel.
The complete state mapping is given in Supplementary Note~\ref{supp:encoding-stability} (Supplementary Table~\ref{supp:tab:timebin_to_polarization}).

Each converter introduces an irreducible \SI{3}{dB} loss: at Alice, half of the optical power is emitted in non-information-carrying time slots produced by the conversion, and at Bob, the 50:50 beam splitter at the polarization-to-time-bin output discards half of the detected events.
These losses are intrinsic to the conversion scheme and are accounted for in the link budget (Supplementary Note~\ref{supp:loss-noise}).

We exploit the converter geometry as a polarization calibration tool.
Choosing different delays $\Delta t_A \neq \Delta t_B$ at Alice and Bob causes any polarization misalignment between Bob's PBS axes and the received states to leak into time slots distinguishable from the information-bearing slots.
Minimizing the leaked signal aligns the PBS axes to the received polarization.
Three fiber-based motorized polarization controllers, inserted before the Bob-side converter, compensate for fiber-induced and atmosphere-induced polarization drift.
Calibration runs use Bob's signal as feedback, so polarization tracking is performed before each session and not in real time during key generation.
The full calibration procedure is described in Supplementary Note~\ref{supp:encoding-stability}.

\subsection{Free-space optical link}
\label{sec:freespace-setup}

\begin{figure}[htbp]
  \centering
  \includegraphics[width=\linewidth,trim={10 190 190 120}, clip]{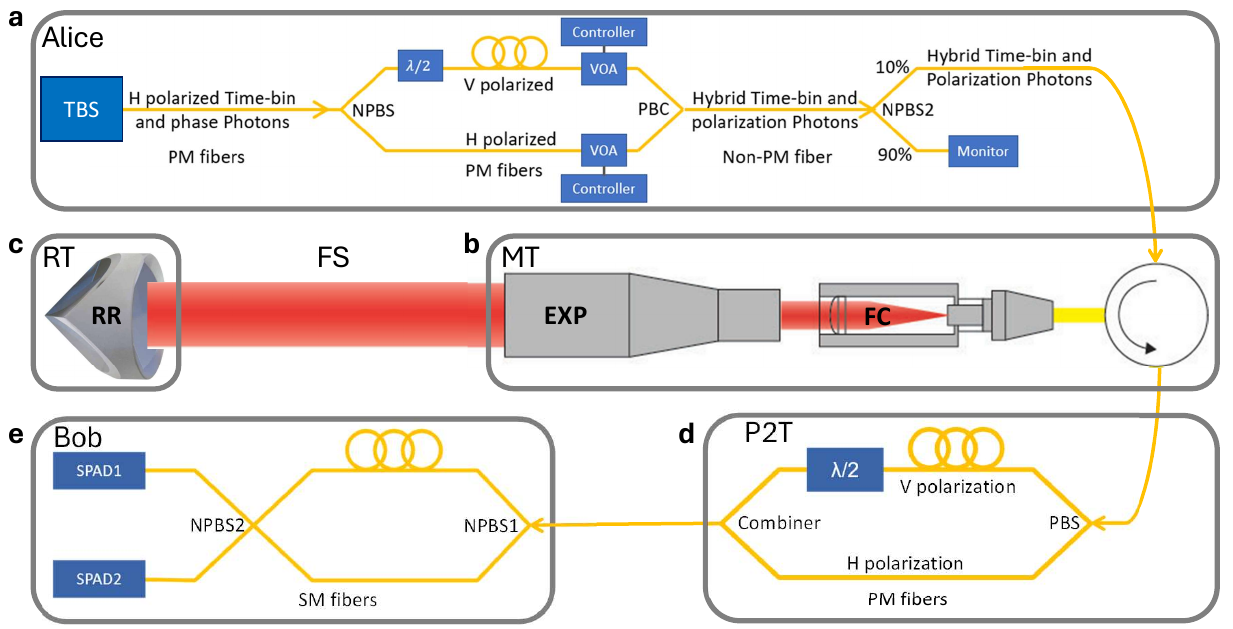}
  \caption{
    \textbf{Hybrid fiber and free-space quantum communication architecture.}
    \textbf{a}~Alice with the T2P converter.
    \textbf{b}~The shared free-space main terminal (MT).
    \textbf{c}~The remote terminal retroreflector (RT).
    \textbf{d}~P2T converter.
    \textbf{e}~Bob's time-bin receiver.
  }
  \label{fig:setup}
\end{figure}

The free-space link uses a folded path in which Alice's transmitter and Bob's receiver share a single fiber-to-free-space port through a single-mode fiber circulator.
Polarization-encoded photons enter port~1, exit port~2 toward the retroreflector, return along the same path, and are routed from port~2 to port~3 toward Bob's detection chain (\cref{fig:setup}\textbf{b},\textbf{c}).
This folded architecture allows hybrid-link operation with a single accurately aligned terminal, removing the need to independently align two terminals.
The same calibration can be reused for retroreflectors at different ranges and locations within the beam's Rayleigh distance, which simplifies deployment.

The shared optical head consists of a fiber collimator, a 90:10 non-polarizing beam splitter (NPBS) that diverts \SI{10}{\percent} of the beam to a SWIR alignment camera, and a beam expander producing a \SI{50.8}{mm} ($2w_0$) collimated beam.
The assembly is mounted on a motorized gimbal with \SI{0.17}{arcsec} angular resolution.
Alignment was performed coarsely on the SWIR camera (with the retroreflector visible in passive light), refined using a CW \SI{1550}{nm} probe injected through the fiber, and finalized by maximizing the power returning at circulator port~3.
Alignment was set manually before each session and was not actively maintained during key generation.

The far end is a passive gold-coated \SI{2}{inch} retroreflector mounted on an adjustable tripod.
Gold coating was chosen to avoid the polarization distortion introduced by uncoated total-internal-reflection retroreflectors~\cite{zhu2014polarization}.
The measured round-trip insertion loss of the free-space module is \SI{6.5}{dB} at \SI{90}{m}, and a detailed loss decomposition is provided in Supplementary Note~\ref{supp:loss-noise}.

\subsection{Atmospheric turbulence characterization}
\label{sec:turbulence-methods}

The refractive-index structure parameter $C_n^2$ was estimated from concurrent meteorological data using Monin-Obukhov surface-layer scaling~\cite{Wyngaard1971} (see Supplementary Note~\ref{supp:turbulence}).
While this simplified parameterization omits wind-shear and stability corrections, the resulting order-of-magnitude estimates capture the pronounced diurnal swing expected in this environment~\cite{andrews2005}.
The estimated $C_n^2$ peaks at ${\approx}\SI{2e-13}{m^{-2/3}}$ in the afternoon, characteristic of strong near-ground daytime turbulence, and falls to ${\approx}\SI{1e-15}{m^{-2/3}}$ at night.

From $C_n^2$, we evaluate the Rytov variance ($\sigma_R^2$), Fried parameter ($r_0$), and aperture-averaged scintillation index ($\sigma_I^2(D)$) using the Gaussian-beam framework of Andrews and Phillips~\cite{andrews2005} for a beam waist $w_0 = \SI{25.4}{mm}$ and wavelength $\lambda = \SI{1550}{nm}$.
These Gaussian corrections are negligible at \SI{90}{m} ($\lesssim$\SI{0.5}{\percent}) but reach $\approx$\SI{22}{\percent} on $r_0$ at \SI{750}{m}, so the plane-wave limit is accurate at the operational \SI{90}{m} distance (see Supplementary Note~\ref{supp:turbulence}).

Although the near-ground $C_n^2$ is strong, the Rytov variance $\sigma_R^2 \propto C_n^2\,L^{11/6}$ stays small because the path is short.
The ratio $D/r_0$, which quantifies the number of turbulence coherence cells across the receiver aperture ($D = \SI{50.8}{mm}$), ranges from ${\approx}0.94$ at peak daytime heating to ${\approx}0.04$ at night, fewer than one cell throughout.
As a result, the short \SI{90}{m} path propagates in the weak-fluctuation regime ($\sigma_R^2 \leq 0.05$) even as $C_n^2$ varied over two orders of magnitude, whereas the same $C_n^2$ reaches a moderate regime ($\sigma_R^2 \approx 2.4$) over the \SI{750}{m} measurements.
We evaluated $\sigma_I^2$ with extended Rytov theory, which departs from the weak-turbulence limit only at \SI{750}{m} ($\sigma_R^2 \gtrsim 1$).
Receiver aperture averaging further suppressed the intensity fluctuations.
At \SI{90}{m} it reduced the point-receiver scintillation index by a factor of ${\approx}30$ (yielding $\sigma_I^2(D) \approx 0.0017$ during peak daytime).

Beam wander and angle-of-arrival (AoA) fluctuations were also evaluated.
Peak daytime beam wander was \SI{1.9}{mm} RMS (${\approx}\SI{8}{\percent}$ of the beam waist) and the maximum RMS AoA reached \SI{12.1}{\micro\radian}, both well within the receiver's field of view.
All turbulence-induced impairments were therefore small at \SI{90}{m}, consistent with the weak propagation regime, and turbulence becomes performance-relevant only at the \SI{750}{m} extension.
Full analytical expressions and the complete diurnal evolution of all parameters are provided in Supplementary Note~\ref{supp:turbulence}.

\section*{Acknowledgments}
The authors thank Raz Bar-On and Amit Erez for their invaluable assistance in the experiments in the early stages of the project.
The authors thank Gal Ben-Shach and Sean Galantzan for their administrative coordination of the project.
K.C. thanks Derek Burrell for his assistance with the turbulence analysis.

\textbf{Funding}
K.C. gratefully acknowledges the Milner Foundation and the VATAT (PBC) Fellowship for Outstanding PhD Students in Data Science.
This work was supported by the Israeli Ministry of Innovation, Science and Technology, Grant No. 1001572598; the Israel Science Foundation (ISF) Excellence Center Grant No. 2312/21 and ISF Grant No. 969/22; and the US-Israel Binational Science Foundation (BSF).

\textbf{Disclosures}
The authors declare no conflicts of interest.

\textbf{Data Availability}
The data that support the plots within this paper and other findings of this study are available from the corresponding author upon reasonable request.

\textbf{Code Availability}
This study did not involve the development of custom code or algorithms central to the physical conclusions.
Standard scripts used solely for data visualization and plot generation are available from the corresponding author upon reasonable request.

\bibliographystyle{naturemag}
\bibliography{references}

\newpage
\appendix

\section{Experimental setup and deployment}
\label{supp:experimental-details}

\subsection{Deployment site and geometry}

The experiment was conducted on the rooftop of the Shenkar building at Tel Aviv University (32.113$^\circ$\,N, 34.804$^\circ$\,E, elevation $\sim$\SI{40}{m} above ground), on 29 March 2026.
Alice and Bob share a single transceiver terminal, and the free-space channel is folded by placing a passive retroreflector as a second terminal (see Fig.~\ref{supp:optical-layout}).
The beam propagates to the retroreflector and back, so all quoted free-space path lengths, losses, and turbulence parameters refer to the total round-trip distance.
For the primary link, the retroreflector stood at a distance of $\sim$\SI{45}{m}, giving a $\sim$\SI{90}{m} round-trip path, with the optical path at a height of $\sim$\SI{1}{m} above the rooftop surface and oriented along 280$^\circ$.
The extended \SI{750}{m} round-trip link reported in Sections~\ref{sec:qkd-results} and~\ref{sec:turbulence} was folded between the Shenkar and Software Engineering buildings, passing $\sim$\SI{20}{m} above urban terrain.

\subsection{Optical layout and components}
\label{supp:optical-layout}

Alice and Bob were built around a pre-commercial decoy-state BB84 system (by HEQA Security) operating at \SI{1550}{nm}.
The system implements the BB84 protocol with decoy states using time-bin encoding, where qubits are encoded in the relative arrival times of photons in two temporal modes (early and late, less than 1\,ns apart).
Two mutually unbiased bases are used: the computational basis $\{|0\rangle, |1\rangle\}$ (early and late time bins) and the superposition basis $\{|+\rangle, |-\rangle\}$ (coherent superpositions of early and late).
The system is specified for single-mode fiber channels with a total loss of up to $\sim$\SI{20}{dB}.

The fiber-coupled output of Alice is converted from time-bin to polarization encoding by a custom fiber-based time-bin to polarization (T2P) converter consisting of an unbalanced Mach-Zehnder interferometer matched to the time-bin delay $\Delta t_A = 800$\,ps, with a $\lambda/2$ phase shift applied to the late-bin arm.
The converter maps each H-polarized time-bin pulse to the corresponding polarization state (Supplementary Table~\ref{supp:tab:timebin_to_polarization}), such that the central output time slot carries the polarization-encoded qubit.
Using the power monitor at the converter output, the variable optical attenuator (VOA) is calibrated to set the mean photon number per pulse $\mu$ at the output.
This conversion introduces an inherent 3\,dB loss: half the output pulse energy falls outside the detection window in the first and last time slots and cannot be compensated by increasing Alice's transmit power.

\begin{figure}[ht]
    \centering
    \includegraphics[width=\linewidth,trim={10 300 0 80}, clip]{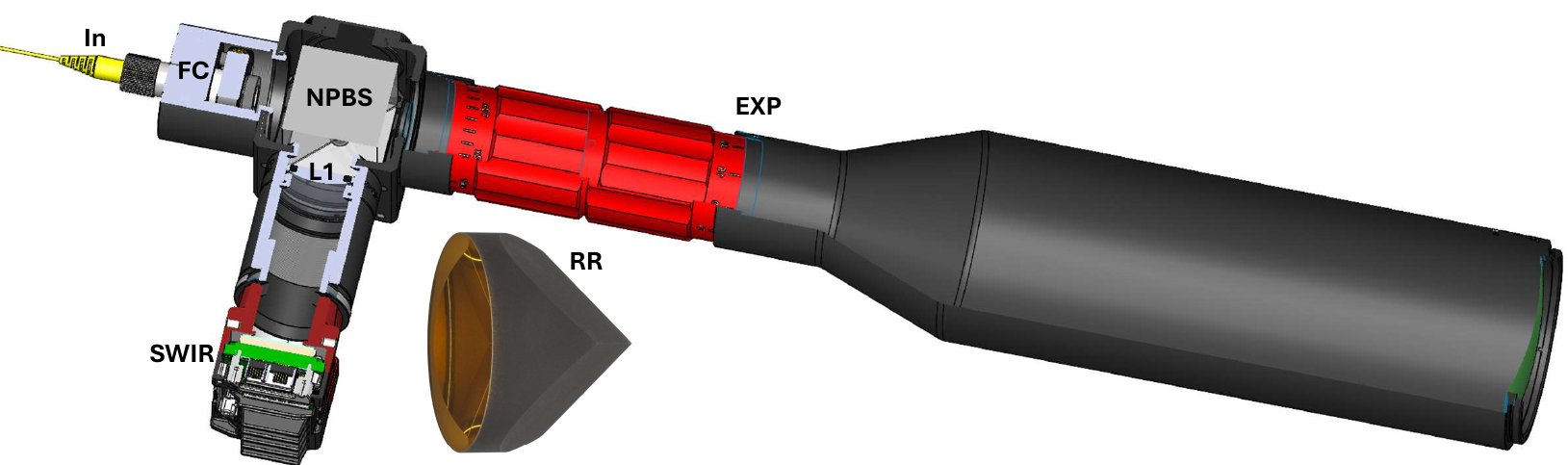}
    \caption{Schematic diagram for the free-space section.}
    \label{fig:freespace_adapter}
\end{figure}

The free-space module (Fig.~\ref{fig:freespace_adapter}) uses a gold-coated 2-inch retroreflector (RR), allowing Alice and Bob to share a single SM fiber terminal connected via a fiber circulator (ports 1$\to$2$\to$3).
The transmitted signal exits port~2 toward a fiber collimator (FC), passes through a 90:10 non-polarizing beam splitter (NPBS), and is expanded to a \SI{2}{inch} diameter by a beam expander (EXP) aimed at the retroreflector (RR) on the far end of the channel.
The reflected beam is recoupled into the same fiber (In) connected to port~2 and routed to port~3 toward the receiver.
A SWIR camera, coupled through the NPBS, allows coarse and fine alignment based on the image and the returning beam spatial-mode.
The entire assembly is mounted on a motorized gimbal with \SI{0.17}{arcsec} angular resolution in azimuth and elevation.
At \SI{90}{m}, the measured round-trip insertion loss is \SI{6.5}{dB} (see Supplementary Note~\ref{supp:loss-noise}).

After the free-space section, three single-mode motorized polarization controllers compensate for polarization drift accumulated in the non-polarization-maintaining fiber components.
These controllers were calibrated at the start of each session by balancing detected photon counts across the time-bin detection slots for each transmitted state (see Supplementary Note~\ref{supp:converter-calibration}), and held fixed during key generation, with the fiber secured to minimize wind-induced movement.

Before Bob, a symmetric polarization to time-bin (P2T) converter maps the received polarization states back to time-bin states: a PBS separates $H$ and $V$ components into independent fiber arms with delay $\Delta t_B \approx 500$\,ps, a $\lambda/2$ waveplate rotates the $V$ component to $H$, and a 50:50 coupler recombines the paths to regenerate the time-bin structure.
The P2T converter delay $\Delta t_B \approx 500$\,ps is set by Bob's downstream time-bin receiver, whose detection interferometer matches this separation.
Since Bob uses only the central polarized pulse, the regenerated time-bin encoding is independent of Alice's encoding delay $\Delta t_A = 800$\,ps.
This step incurs a 3\,dB loss since only one coupler output port is detected.
The discarded port carries no security penalty even if it is sniffed.
The light it emits is a passive, deterministic linear-optical combination of the photons entering Bob's converter from the quantum channel, so any information an eavesdropper could extract there she could already obtain by tapping the channel upstream, an attack the BB84 decoy-state security analysis already accounts for by granting her full control of the channel~\cite{scarani2009}.
Any such leakage, including from multi-photon pulses, is bounded by the same analysis and removed by privacy amplification, so the distilled key remains information-theoretically secure.

A photograph of the deployed setup is shown in Fig.~\ref{fig:main}\textbf{b}, and a schematic of the Alice and Bob terminals is shown in Fig.~\ref{fig:setup}.



\section{Loss budget and noise}
\label{supp:loss-noise}

\subsection{Free-space module loss decomposition}
\label{supp:loss-decomposition}

The \SI{6.5}{dB} round-trip insertion loss of the free-space module at \SI{90}{m} (referenced in Methods, Section~\ref{sec:freespace-setup}) decomposes as: \SI{0.94}{dB} for the circulator port~1$\to$2 transition, $2\times\SI{0.46}{dB}$ for the forward and return passes through the 90:10 NPBS, \SI{1.43}{dB} for the port~2$\to$3 transition, and the remaining \SI{3.2}{dB} for free-space propagation and recoupling.
Bench measurements isolate the recoupling component as the dominant contribution to the \SI{3.2}{dB} term: with the retroreflector connected directly to the optical head (no free-space propagation, no beam expansion), the same loss is observed.
In this bench characterization, substituting a total internal reflection (TIR) test retroreflector with a \SI{1}{inch} cat-eye retroreflector reduced the free-space loss by \SI{1.8}{dB}.
The remaining \SI{1.4}{dB} is attributed to imperfect optical alignment.
The deployed link used the gold-coated \SI{2}{inch} retroreflector throughout (Section~\ref{sec:freespace-setup}).

Since the Rayleigh range of the 2-inch-diameter beam ($w_0 = \SI{25.4}{mm}$, $\lambda = \SI{1550}{nm}$) is
\begin{equation}
    z_R = \frac{\pi w_0^2}{\lambda} \approx \SI{1.3}{km},
    \label{eq:Rayleigh_range}
\end{equation}
the free-space diffraction loss is negligible at distances $\ll z_R$.
At \SI{90}{m} (${\approx}0.07\,z_R$), no excess path loss beyond the calibrated module baseline was observed, consistent with the link lying well within the collimated regime.
At \SI{750}{m}, the measured round-trip path loss was \SI{10.3}{dB}, i.e.\ approximately \SI{3.8}{dB} above the \SI{90}{m} baseline.
This excess is consistent with an effective beam waist radius $w_0 \approx \SI{13.3}{mm}$ ($z_R \approx \SI{360}{m}$), placing the \SI{750}{m} link at ${\approx}2\,z_R$, which we attribute to imperfect beam-expander focus calibration in the field (a reduced effective waist adds diffraction and recoupling loss), compounded by turbulence-induced coupling loss.

\subsection{End-to-end loss table}

Table~\ref{supp:tab:loss-budget} lists each element from Alice's source to Bob's detectors and its measured or estimated insertion loss for the \SI{90}{m} link.
Because Alice and Bob share a single fiber terminal (retroreflector geometry), the free-space module loss is a single round-trip quantity.

\begin{table}[htbp]
\centering
\small
\caption{
  End-to-end optical loss budget for the \SI{90}{m} free-space QKD link.
  Values marked with $^\dagger$ are directly measured, and the rest are estimated.
}
\begin{tabular}{lcc}
\toprule
Component & Loss (dB) & Notes \\
\midrule
\multicolumn{3}{l}{\textit{Alice side}} \\
T2P converter                      & $3.0^\dagger$ & Sec.~\ref{sec:converter}, early/late pulse loss \\
\multicolumn{3}{l}{\textit{Free-space module (round-trip, port~1 $\to$ port~3), see~\ref{supp:loss-decomposition}}} \\
Circulator (port~1$\to$2)          & $0.94^\dagger$ & \\
NPBS (forward, 90:10)              & $0.46^\dagger$ & \\
Fiber collimator + beam expander   & 3.2            & Propagation and recoupling \\
Free-space path (\SI{90}{m})       & ${\approx}0$   & $90\,\text{m} \ll z_R$ (Eq.~\ref{eq:Rayleigh_range}) \\
Gold retroreflector                & $<0.1$         & ${>}97\%$ at \SI{1550}{nm} \\
NPBS (return, 90:10)               & $0.46^\dagger$ & \\
Circulator (port~2$\to$3)          & $1.43^\dagger$ & \\
\textit{FSM subtotal}              & $6.5^\dagger$  & Measured end-to-end \\
\multicolumn{3}{l}{\textit{Bob side}} \\
P2T converter                      & $3.0^\dagger$ & Sec.~\ref{sec:converter}. BS recombination loss \\
Fiber patch cords \& connectors    & ${\sim}0.5$--$1.0$ & Estimated \\
\midrule
\textbf{Optical total (to detector input)} & ${\approx}\mathbf{13}$ & Converters + FSM + interconnects \\
\bottomrule
\end{tabular}
\label{supp:tab:loss-budget}
\end{table}

\section{Encoding stability and converter calibration}
\label{supp:encoding-stability}

\subsection{T2P converter calibration}
\label{supp:converter-calibration}

We exploit a deliberate mismatch between the converter delays of Alice and Bob ($\Delta t_A \neq \Delta t_B$) as a polarization calibration tool.
When Bob's PBS axes are misaligned with the polarization states received from the channel, a fraction of the $X$-basis pulses passes through both arms of the P2T converter and produces a coherent time-bin signal with separation $\Delta t_B$ in slots distinguishable from the information-carrying slots.
By maximizing the signal in the correct slots and minimizing the leaked signal in the others, the P2T converter is aligned to the incoming polarization basis.
The mismatch $\Delta t_A \neq \Delta t_B$ also confirms that the two converters operate independently, so that the same hardware can be used in either direction.
Three motorized fiber polarization controllers before the P2T converter compensate for fiber- and atmosphere-induced drift.
Because their feedback uses Bob's detected signal, calibration is performed before each session and not re-applied during key generation.

The complete state mapping of the Alice-side converter is given in Table~\ref{supp:tab:timebin_to_polarization}.
The qubit is carried in the central output slot $t_1$.
The side-slot pulses at $t_0$ and $t_2$ account for the inherent 3\,dB loss.

\begin{table}[h]
\centering
\renewcommand{\arraystretch}{1.4}
\caption{Time-bin to polarization state mapping at the Alice converter output.
Pulses are labelled by time slot.
$\tau_{800}$ and $\tau_{1600}$ are the single and double delay-line intervals.
The qubit information is encoded in the polarization state at the central slot $t_1$.
Probability entries sum to~1 per input state, and the side-slot pulses represent the 3\,dB loss.}
\label{supp:tab:timebin_to_polarization}
\small
\begin{tabular}{c ccc ccc ccc}
\toprule
& \multicolumn{3}{c}{$t_0$}
& \multicolumn{3}{c}{$t_1 = t_0 + \tau_{800}$}
& \multicolumn{3}{c}{$t_2 = t_0 + \tau_{1600}$} \\
\cmidrule(lr){2-4}\cmidrule(lr){5-7}\cmidrule(lr){8-10}
Input state & Prob. & & Polarization & Prob. & & Polarization & Prob. & & Polarization \\
\midrule
$|0\rangle$ & $\tfrac{1}{2}$ && $|H\rangle$ & $\tfrac{1}{2}$ && $|V\rangle$                         & ---            && --- \\
$|1\rangle$ & ---            && ---         & $\tfrac{1}{2}$ && $|H\rangle$                         & $\tfrac{1}{2}$ && $|V\rangle$ \\
$|+\rangle$ & $\tfrac{1}{4}$ && $|H\rangle$ & $\tfrac{1}{2}$ && $|H\rangle + |V\rangle \equiv |D\rangle$ & $\tfrac{1}{4}$ && $|V\rangle$ \\
$|-\rangle$ & $\tfrac{1}{4}$ && $|H\rangle$ & $\tfrac{1}{2}$ && $|H\rangle - |V\rangle \equiv |A\rangle$ & $\tfrac{1}{4}$ && $-|V\rangle$ \\
\bottomrule
\end{tabular}
\end{table}

\subsection{Fiber-only stability experiment}
\label{app:fiberQBEReval}

To isolate the contribution of the fiber channel to the overall quantum bit error rate, we performed a dedicated calibration measurement using time-bin BB84 states transmitted through fiber with a variable optical attenuator simulating propagation distances of \SIrange{7.8}{111.1}{km}.
Table~\ref{tab:fiber-qber} summarizes the results.
The measured QBER remains essentially flat at $\sim$1.45\% across all attenuation levels, confirming that this error floor is an intrinsic property of the transmitter-receiver hardware rather than a loss-dependent effect.
The $X$-basis QBER ($\sim$2.3\%) is consistently higher than the $Z$-basis ($\sim$0.7\%), as expected for time-bin encoding where the superposition basis requires interferometric detection and is therefore more sensitive to phase drift.

\begin{table}[htbp]
\centering
\caption{Fiber-only QBER measured with time-bin BB84 states at varying channel attenuation.}
\label{tab:fiber-qber}
\small
\begin{tabular}{ccccc}
\toprule
Attenuation & Equiv.\ fiber & \multicolumn{3}{c}{QBER (\%)} \\
\cmidrule(lr){3-5}
(dB) & (km) & Overall & $Z$-basis & $X$-basis \\
\midrule
1.4  &   7.8 & 1.44 & 0.63 & 2.32 \\
4.0  &  22.2 & 1.33 & 0.61 & 2.11 \\
8.0  &  44.4 & 1.43 & 0.66 & 2.27 \\
12.0 &  66.7 & 1.47 & 0.66 & 2.35 \\
16.0 &  88.9 & 1.51 & 0.75 & 2.31 \\
20.0 & 111.1 & 1.55 & 0.88 & 2.27 \\
\midrule
\multicolumn{2}{c}{Average} & 1.45 & 0.70 & 2.27 \\
\bottomrule
\end{tabular}
\end{table}

\subsection{Free-space polarization stability evaluation}
\label{app:fsQBEReval}
As mentioned, polarization-based encoding is preferred in free space.
To confirm this, we performed a separate experiment in which we transmitted the four BB84 states ($0\degree$, $90\degree$, $\pm45\degree$) at \SI{1550}{nm} through a retroreflector link and recorded the received azimuth with a polarimeter at six daytime sites over round-trip ranges of \SIrange{40}{200}{m} (Table~\ref{tab:pol-stability}).
Note that this was a separate experiment, performed at a different time and under different conditions (Table~\ref{tab:pol-sites}).

\begin{table}[htbp]
\centering
\caption{
  Polarization measurement site conditions.
  Range is the round-trip propagation distance through the retroreflector.
}
\label{tab:pol-sites}
\small
\begin{tabular}{lcccccccc}
\toprule
Site & Range & Incl. & Azimuth & Alt. & Humidity & Pressure & Wind & Time \\
     & (m)   & (\degree) & (\degree) & (m) & (\%) & (mbar) & (km/h) & \\
\midrule
1F &  40 &   0    & 270 &  58 & 61 & 1006.5 & 15 & 16:20 \\
1A & 100 &   0    & 090 & 141 & 15 & 995.9  & 23 & 10:56 \\
1B & 104 & $-$15  & 310 &  53 & 60 & 1008   & 10 & 14:00 \\
1C & 150 & $-$17  & 300 &  53 & 60 & 1008   & 15 & 14:30 \\
1D & 180 & $-$45  & 030 &  15 & 61 & 1009.5 & 15 & 15:15 \\
1E & 200 & $-$40  & 026 &  15 & 61 & 1009.5 & 15 & 15:30 \\
\bottomrule
\end{tabular}
\end{table}

From the measured azimuthal deviation $\delta$, the implied QBER of a projective BB84 measurement is $\sin^2\delta$ (Malus's law).
The mean deviation stayed below $2.7\degree$ ($<0.3\%$), and the worst-case instantaneous deviation, $13.4\degree$, gives $5.3\%$, below the $11\%$ threshold.
These results are consistent with our main experiment and indicate the robustness of free-space polarization encoding at \SI{1550}{nm}.

\begin{table}[htbp]
\centering
\caption{Free-space polarization stability versus round-trip range (worst case over the four polarization states). $\bar\delta$ stands for mean azimuthal deviation. $\delta_{\max}$ stands for maximum deviation. QBER $=\sin^2\delta_{\max}$. DOP stands for minimum degree of polarization.}
\label{tab:pol-stability}
\small
\begin{tabular}{lccccc}
\toprule
Site & Range (m) & $\bar\delta$ (\degree) & $\delta_{\max}$ (\degree) & QBER (\%) & DOP (\%) \\
\midrule
1F &  40 & 1.5 &  4.6 & 0.6 & 96.4 \\
1A & 100 & 1.4 & 10.3 & 3.2 & 96.9 \\
1B & 104 & 2.7 & 11.3 & 3.8 & 98.1 \\
1C & 150 & 1.5 &  9.5 & 2.7 & 96.0 \\
1D & 180 & 1.8 & 13.4 & 5.3 & 95.5 \\
1E & 200 & 1.7 & 10.0 & 3.0 & 95.6 \\
\bottomrule
\end{tabular}
\end{table}


\section{Extended atmospheric turbulence analysis}
\label{supp:turbulence}

\subsection{Recorded conditions and \texorpdfstring{$C_n^2$} {Cn2} estimation}
The link was operated on 29 March 2026.
No direct $C_n^2$ measurement was available, so we estimated it from a Monin-Obukhov surface-layer model~\cite{Wyngaard1971,Frederickson2000}.
Optical turbulence is set by the temperature structure parameter
\begin{equation}
  C_T^2 = 4.9\,\theta_*^2\, z^{-2/3},
  \qquad
  C_n^2 = \left(\frac{79 \times 10^{-6}\,P}{T^2}\right)^{\!2}\! C_T^2,
  \label{eq:cn2}
\end{equation}
where $\theta_*$ is the surface-layer temperature scale set by the surface heat flux and friction velocity~\cite{Wyngaard1971}, $z = \SI{1.5}{m}$ is the beam height, $P$ in mbar and $T$ in K.
Table~\ref{tab:turbenv} lists the recorded hourly rooftop conditions and the estimated $C_n^2$.

For daytime convective surface layers $\theta_* \approx 0.1$--$\SI{0.5}{K}$ \cite{Wyngaard1971,Frederickson2000}.
We adopt $\theta_* \approx \SI{0.25}{K}$ at peak heating, yielding $C_n^2 \approx \SI{2e-13}{m^{-2/3}}$, the upper end of reported near-ground daytime values.

\begin{table}[htbp]
\label{tab:turbenv}
\centering
{\caption{Recorded rooftop conditions (29 March 2026) and the estimated $C_n^2$.}}
\small
\begin{tabular}{lccccc}
\toprule
{Hour} & {$T$ (\degree C)} & {RH (\%)} & {Dew pt.\ (\degree C)} & {$U$ (m/s)} & {$C_n^2$ ($10^{-13}$\,m$^{-2/3}$)} \\
\midrule
{15:00} & {18} & {65}   & {12}  & {3.1} & {2.08}   \\
{16:00} & {18} & {65}   & {12}  & {3.1} & {1.60}   \\
{17:00} & {18} & {62}   & {11}  & {2.8} & {0.814}  \\
{18:00} & {18} & {62}   & {10}  & {2.5} & {0.293}  \\
{19:00} & {17} & {64}   & {10}  & {2.2} & {0.0845} \\
{20:00} & {16} & {64}   & {9.5} & {1.7} & {0.0335} \\
{21:00} & {16} & {64.5} & {9}   & {1.4} & {0.0121} \\
{22:00} & {16} & {65.5} & {9}   & {1.5} & {0.0121} \\
\bottomrule
\end{tabular}
\end{table}

\subsection{Gaussian-beam corrections to the plane-wave expressions}
The Gaussian-beam Fried parameter and Rytov variance follow Andrews and Phillips~\cite{andrews2005}:
\begin{equation}
  r_0 = \left[ 0.423\, k^2 C_n^2 L
  \left(1 + 1.5\,\Theta_0^2 (1-\Lambda)^2 \right)^{-5/6}\right]^{-3/5},
  \label{eq:r0-gauss}
\end{equation}
\begin{equation}
  \sigma_R^2 = 1.23\, C_n^2\, k^{7/6}\, L^{11/6}
  \left[ 1 + 0.18\, \Theta_0^{12/5}(1-\Lambda)^{12/5} \right]^{-7/12},
  \label{eq:rytov-gauss}
\end{equation}
where $\Theta_0 = L/z_R$, $z_R = \pi w_0^2/\lambda$ is the Rayleigh range, and $\Lambda = 0$ for a collimated beam.
For our beam ($w_0 = \SI{25.4}{mm}$, the $1/e^2$ radius of the 2-inch collimated beam, $\lambda = \SI{1550}{nm}$, $z_R \approx \SI{1.3}{km}$), $\Theta_0 = 0.069$ at \SI{90}{m} and $0.574$ at \SI{750}{m}.
The corrections are negligible at \SI{90}{m} ($\lesssim\!\SI{0.5}{\percent}$) and reach $\approx\!\SI{22}{\percent}$ on $r_0$ at \SI{750}{m}, so the plane-wave limit is accurate at the operational distance.

\subsection{Strong-turbulence scintillation index}
For weak turbulence $\sigma_I^2 \approx \sigma_R^2$.
More generally we use the extended Rytov expressions of Andrews and Phillips~\cite{andrews2005}, which split the log-irradiance variance into large- and small-scale parts:
\begin{align}
  \sigma_{\ln x}^2 &= \frac{0.49\,\sigma_R^2}
    {\left(1 + 1.11\,\sigma_R^{12/5}\right)^{7/6}}, \qquad
  \sigma_{\ln y}^2 = \frac{0.51\,\sigma_R^2}
    {\left(1 + 0.69\,\sigma_R^{12/5}\right)^{5/6}}, \label{eq:strong-components} \\
  \sigma_I^2 &= \exp(\sigma_{\ln x}^2 + \sigma_{\ln y}^2) - 1, \label{eq:strong-scint}
\end{align}
with $\sigma_R = (\sigma_R^2)^{1/2}$, so that $\sigma_R^{12/5} = (\sigma_R^2)^{6/5}$.
These reduce to $\sigma_I^2 \approx \sigma_R^2$ for $\sigma_R^2 \ll 1$ and reproduce the saturation of scintillation, which peaks near $\sigma_R^2 \approx 10$ and then relaxes toward unity.
Over our links $\sigma_R^2 \lesssim 2.4$ (Table~\ref{tab:turb-750m-full}).

\subsection{Aperture averaging, beam wander, and angle of arrival}
The aperture-averaged scintillation index~\cite{andrews2005} embeds the aperture in both the large and small scale terms and is valid in all regimes:
\begin{align}
  {\sigma_I^2(D)} &{= \exp\!\left[\sigma_{\ln X}^2(D)
       + \sigma_{\ln Y}^2(D)\right] - 1,} \nonumber \\
  {\sigma_{\ln X}^2(D)} &{= \frac{0.49\,\sigma_R^2}
       {\left(1 + 0.65\,d^2 + 1.11\,\sigma_R^{12/5}\right)^{7/6}},} \nonumber \\
  {\sigma_{\ln Y}^2(D)} &{= \frac{0.51\,\sigma_R^2
       \left(1 + 0.69\,\sigma_R^{12/5}\right)^{-5/6}}
       {1 + 0.90\,d^2 + 0.62\,d^2\,\sigma_R^{12/5}},}
  \label{eq:aperture-avg}
\end{align}
with $d^2 = kD^2/4L = (D/2 r_F)^2$ and Fresnel scale $r_F = \sqrt{\lambda L /(2\pi)}$.
For our \SI{50.8}{mm} aperture, $d^2 = 29$ ($r_F = \SI{4.7}{mm}$) at \SI{90}{m} and $d^2 = 3.5$ ($r_F = \SI{13.6}{mm}$) at \SI{750}{m}, reducing the point-receiver $\sigma_I^2$ by $\approx\!30\times$ ($0.050 \to 0.0017$ at peak) and $\approx\!5\times$ ($1.04 \to 0.20$), respectively.
Beam wander and angle-of-arrival (AoA) variances are
\begin{equation}
  \langle r_c^2 \rangle = 7.25\, C_n^2\, L^3\, w_0^{-1/3},
  \qquad
  \langle \theta^2 \rangle = 2.91\, C_n^2\, L\, D^{-1/3}.
  \label{eq:beam-wander-aoa}
\end{equation}
For $w_0 = \SI{25.4}{mm}$, beam wander at \SI{90}{m} ranges from \SI{0.15}{mm} (night) to \SI{1.9}{mm} RMS (day), $\lesssim\!8\%$ of the beam waist and well within the receiver aperture, and AoA from {\SIrange{0.9}{12.1}{\micro\radian}}.

\subsection{Full turbulence tables}
Tables~\ref{tab:turb-90m-full} and~\ref{tab:turb-750m-full} list the complete hourly turbulence
parameters at both distances.
\begin{table}[htbp]
\centering
\caption{Complete turbulence parameters at \SI{90}{m} (Gaussian beam,
$w_0 = \SI{25.4}{mm}$).}
\label{tab:turb-90m-full}
\small
\begin{tabular}{lccccccc}
\toprule
Hour & $C_n^2$ & $\sigma_R^2$ & $r_0$ & $\sigma_I^2$ & $\sigma_I^2$(ap) & BW RMS & AoA RMS \\
     & {($10^{-13}$\,m$^{-2/3}$)} & & (cm) & & {($10^{-3}$)} & (mm) & ($\mu$rad) \\
\midrule
15:00 & {2.08}  & {0.0502}   & {5.40}  & {0.0502}   & {1.66}   & {1.94} & {12.1} \\
16:00 & {1.60}  & {0.0384}   & {6.34}  & {0.0385}   & {1.28}   & {1.69} & {10.6} \\
17:00 & {0.814} & {0.0196}   & {9.49}  & {0.0196}   & {0.658}  & {1.21} & {7.59} \\
18:00 & {0.293} & {0.00706}  & {17.52} & {0.00707}  & {0.238}  & {0.73} & {4.55} \\
19:00 & {0.0845}& {0.00204}  & {36.95} & {0.00204}  & {0.0687} & {0.39} & {2.44} \\
20:00 & {0.0335}& {0.000806} & {64.41} & {0.000806} & {0.0272} & {0.25} & {1.54} \\
21:00 & {0.0121}& {0.000290} & {118.9} & {0.000290} & {0.0098} & {0.15} & {0.92} \\
22:00 & {0.0121}& {0.000290} & {118.9} & {0.000290} & {0.0098} & {0.15} & {0.92} \\
\bottomrule
\end{tabular}
\end{table}
\begin{table}[htbp]
\centering
\caption{Complete turbulence parameters at \SI{750}{m} (Gaussian beam,
$w_0 = \SI{25.4}{mm}${, the daytime link reaching the moderate regime,
$\sigma_R^2 \lesssim 2.4$}).}
\label{tab:turb-750m-full}
\small
\begin{tabular}{lccccccc}
\toprule
Hour & $C_n^2$ & $\sigma_R^2$ & $r_0$ & $\sigma_I^2$ & $\sigma_I^2$(ap) & BW RMS & AoA RMS \\
     & {($10^{-13}$\,m$^{-2/3}$)} & & (cm) & & & (mm) & ($\mu$rad) \\
\midrule
15:00 & {2.08}  & {2.383}  & {1.84}  & {1.044}  & {0.199}   & {46.6} & {35.0} \\
16:00 & {1.60}  & {1.825}  & {2.16}  & {0.952}  & {0.189}   & {40.7} & {30.7} \\
17:00 & {0.814} & {0.931}  & {3.24}  & {0.676}  & {0.145}   & {29.1} & {21.9} \\
18:00 & {0.293} & {0.335}  & {5.98}  & {0.309}  & {0.0711}  & {17.5} & {13.1} \\
19:00 & {0.0845}& {0.0967} & {12.61} & {0.0959} & {0.0230}  & {9.38} & {7.06} \\
20:00 & {0.0335}& {0.0383} & {21.98} & {0.0383} & {0.00934} & {5.90} & {4.44} \\
21:00 & {0.0121}& {0.0138} & {40.57} & {0.0138} & {0.00339} & {3.54} & {2.66} \\
22:00 & {0.0121}& {0.0138} & {40.57} & {0.0138} & {0.00339} & {3.54} & {2.66} \\
\bottomrule
\end{tabular}
\end{table}

\section{Security analysis and post-processing}
\label{supp:post-processing}

\subsection{Decoy-state secure-key analysis}
The fraction of disagreeing bits in the sifted key defines the quantum bit error rate,
\begin{equation}
  \mathrm{QBER} = \frac{n_{\mathrm{error}}}{n_{\mathrm{sifted}}},
  \label{eq:qber}
\end{equation}
reported separately for the signal and decoy intensities.
The decoy-state analysis follows the Gottesman--Lo--L\"utkenhaus--Preskill (GLLP) framework\cite{Gottesman2004}, bounding the single-photon gain $Q_1$ and phase-error rate $e_1$ to give the secure key rate as
\begin{equation}
  R_{\mathrm{secure}} \geq q \left\{ Q_1 \left[ 1 - H(e_1) \right]
  - Q_\mu\, f_{\mathrm{EC}}\, H(E_\mu) \right\},
  \label{eq:secure-rate}
\end{equation}
where $q = 1/2$ is the basis-sifting factor, $Q_\mu$ is the overall gain, $E_\mu$ is the measured QBER, $f_{\mathrm{EC}} \geq 1$ is the error-correction efficiency, and $H(\cdot)$ is the binary Shannon entropy.
The single-photon gain lower bound $Q_1^{\mathrm{LB}}$ and phase-error rate upper bound $e_1^{\mathrm{UB}}$ are estimated from the measured signal and decoy gains and error rates following the standard analysis of Ma \textit{et~al.}~\cite{ma2005}.
As a receiver figure of merit, the interferometric visibility of Bob's Mach-Zehnder interferometer, $V = (N_{\max} - N_{\min}) / (N_{\max} + N_{\min})$ with $N$ the photon counts in the constructive and destructive ports, is monitored throughout each session.
\subsection{Error correction and privacy amplification}
Sifted-key disagreements arising from channel noise and detector imperfections are reconciled with the \textsc{Cascade} protocol~\cite{brassard1993secret}, which exchanges parity information over an authenticated public channel.
Across our sessions, the error-correction efficiency was $f_{\mathrm{EC}} \approx 1.1$.
Privacy amplification then applies a universal hash function that removes both the information disclosed during reconciliation and the eavesdropper's bound on the single-photon contribution, yielding the final secure key.
These two effects are, respectively, the $Q_\mu f_{\mathrm{EC}} H(E_\mu)$ and $Q_1[1 - H(e_1)]$ terms of Eq.~\ref{eq:secure-rate}, and together they set the gap between the sifted and secure key rates.



\end{document}